\documentclass[12pt]{article}
\input axodraw.sty
\newcommand{\Tint}[1]{{\hbox{$\sum$}\!\!\!\!\!\!\int}_{\!\!\!\!#1}}
\newcommand{\la}[1]{\label{#1}}
\newcommand{\be}{\begin{equation}}
\newcommand{\ee}{\end{equation}}
\newcommand{\ba}{\begin{eqnarray}}
\newcommand{\ea}{\end{eqnarray}}
\newcommand{\bi}{\begin{itemize}}
\newcommand{\ei}{\end{itemize}}

\newcommand{\nr}[1]{(\ref{#1})}

\newcommand{\nn}{\nonumber \\}
\newcommand{\fr}[2]{{\frac{#1}{#2}}}

\renewcommand{\vec}[1]{{\bf #1}}
\newcommand{\eq}{Eq.~}

\def\lsi{\raise0.3ex\hbox{$<$\kern-0.75em\raise-1.1ex\hbox{$\sim$}}}
\def\gsi{\raise0.3ex\hbox{$>$\kern-0.75em\raise-1.1ex\hbox{$\sim$}}}
\newcommand{\lsim}{\mathop{\lsi}}

\begin{document}

\begin{flushright}
CERN-TH/99-353\\
UNIL-IPT/99-6\\
hep-ph/9911473\\
January 2000
\end{flushright}
\begin{centering}

{\bf A REMARK ON SPHALERON ERASURE\\ OF BARYON ASYMMETRY}
\vspace{0.8cm}

M. Laine$^{\rm a,b,}$\footnote{mikko.laine@cern.ch}, 
M. Shaposhnikov$^{\rm c,}$\footnote{mikhail.shaposhnikov@ipt.unil.ch} \\

\vspace{0.3cm}
{\em $^{\rm a}$Theory Division, CERN, CH-1211 Geneva 23,
Switzerland\\}
\vspace{0.3cm}
{\em $^{\rm b}$Dept.\ of Physics,
P.O.Box 9, FIN-00014 Univ.\ of Helsinki, Finland\\}
\vspace{0.3cm}
{\em $^{\rm c}$Institute for Theoretical Physics, University of  
Lausanne,\\ 
BSP-Dorigny, CH-1015 Lausanne, Switzerland}

\vspace{0.7cm}
{\bf Abstract}

\end{centering}

\vspace{0.3cm}\noindent 

We complete an existing result for how the baryon asymmetry 
left over after a period of full thermal equilibrium depends
on different lepton asymmetries. 





\paragraph{Introduction.}

Baryon plus lepton number ($B+L$) is violated by anomalous 
electroweak processes, and at the high temperatures appearing in the
Early Universe these processes are fast enough to be  in thermal
equilibrium~\cite{krs1}. This leaves only three conserved global
charges, 
\be
X_i = \frac{1}{n_F} B - L_i,
\ee
where $i=1,...,n_F$, and $n_F$ denotes the number of
generations\footnote{In extensions of  the Minimal Standard Model
even these charges may be violated,  for instance due to flavour
non-diagonal slepton masses or  Majorana-type right-handed neutrino
masses, but we assume here that the $X_i$ are strictly conserved.}.

For cosmological applications it is important to know what is the
value of the baryon number $B$ in an equilibrium system with given
$X_i$ (see, e.g., a recent work~\cite{mmr}). This problem has been
addressed in a number of papers~\cite{krs2}--\cite{ks96}, and different
answers to the same questions have been given. The
confusing points were related to accounting for the scalar degrees of
freedom and to fixing the boundary conditions for the gauge charges
(electric charge versus hypercharge in the Higgs phase of the
theory). 

In the limit of small Yukawa couplings the baryon number can be
written in the form 
\be
B = f_0\sum_i X_i + \sum_i X_i 
\biggl(f_1 \frac{m_i^2}{T^2} + f_2 h_i^2\biggr), \la{form}
\ee
where $m_i = h_i\phi/\sqrt{2}$ are the lepton masses,
$h_i$ are the lepton Yukawa couplings, $\phi$ is
the expectation value of the Higgs field, $T$ is 
the temperature, and $\sum_i X_i= B-L$. 
The $f_i$ are some functions of $T$, $\phi$,
the number of fermionic flavours $n_F$, and 
the number of Higgs scalars $n_S$~\footnote{The 
definitions of $f_1,f_2$ are actually not completely unique, since
a term proportional to $\phi^2$ in $f_2$ can equivalently
be presented as a contribution to $f_1$. We make here
a certain division based on the way in which the
two terms arise in our computation.}. 
The function $f_0$ was computed correctly 
in the symmetric phase  
of the EW theory in~\cite{ks88}, 
and in the Higgs phase in~\cite{ks96}; the
limit $\phi\gg T$ of $f_0$ coincides with the results of 
refs.~\cite{ht,nb,d,dr}. 

The situation with $f_1$ and $f_2$ is more obscure. In 
ref.~\cite{ks88} it was claimed that 
$f_1 = (4/13 \pi^2)A$, $f_2 = (1/13 \pi^2)A$ with
$A \simeq 1$ for both symmetric and the Higgs phases provided the
temperature is much larger than the vector boson masses
(however, no explicit expression for $A$ was presented). 
This result was stated to be wrong in~\cite{dr} based on 
a reasoning related to the boundary conditions, and
another form for $f_1$ was derived. In~\cite{dr} it was said, 
moreover, that in the symmetric phase $B=0$ if $B-L=0$, 
meaning effectively that $f_2=0$. In~\cite{dko}, 
on the contrary, a specific non-vanishing expression
was derived for $f_2$ in the limit $\phi\ll T$.

The question of how to really carry out 
the computation for generic $\phi,T$ was finally
clarified in~\cite{ks96}, but no estimates were
presented for $f_1$ and $f_2$, as 
only $f_0$ was computed. We feel that the correct results
for $f_1$ and $f_2$, valid both in the symmetric and the Higgs
phases, should  appear in the literature at last, and this is the
aim of the present note. 

\paragraph{Method.}

Let us briefly recall the method of computing the  baryon
number~\cite{ks96}. We have $n_F$ conserved global charges $X_i$,
with which we can associate chemical potentials.  However, it is more
convenient to introduce first the  $n_F+1$ chemical potentials
$\mu_B,\mu_{L_i}$, and impose the constraint following from the
sphaleron processes,
\be
n_F \mu_B + \sum_{i=1}^{n_F} \mu_{L_i} =0
\la{sphconstraint}
\ee
only later on. We now have to compute the effective potential 
$V(\phi,A_0^a,B_0; \mu_B,\mu_{L_i},T)$, where $\phi$ is the Higgs
expectation value, and $A_0^a,B_0$ are the temporal components of the
SU(2) and U(1) gauge fields.  The fields $A_0^a,B_0$ have to be
included, since the  chemical potentials break Lorentz invariance 
and induce expectation values for them. The minimization ${\partial
V}/{\partial A_0^a}= {\partial V}/{\partial B_0}=0$ corresponds to
the neutrality of the system with respect to gauge charges. From the
gauge-invariant value of the effective potential at the  minimum, we
get $B = -{\partial V}/{\partial \mu_B},  L_i = -{\partial
V}/{\partial \mu_{L_i}}$ (these are really the baryon and lepton
numbers per unit volume). Utilising \eq\nr{sphconstraint}, we can
finally eliminate $\mu_B,\mu_{L_i}$ to obtain an expression of the
desired form $B = f(X_i)$.

We will work at high temperatures, assuming $\mu_B,\mu_{L_i}\ll T$
and $\phi\lsim (\mbox{a few})\times T$.
The expectation values of $A_0^a,B_0$ are proportional to the $\mu$'s, 
so it is sufficient to keep terms up to quadratic order in 
$A_0^a,B_0,\mu_B,\mu_{L_i}$.  It is sufficient to choose
only $A_0^3$ non-zero. The bosonic degrees of freedom (Higgses, 
gauge fields) only contribute to terms quadratic in $A_0^3,B_0$, 
while $\mu_B,\mu_{L_i}$ come from the fermionic contributions. 
We denote the fermionic terms involving 
$A_0^a,B_0,\mu_B,\mu_{L_i}$ by

\parbox[c]{2.4cm}{
\begin{picture}(140,40)(0,0)

\SetWidth{1.5}
\SetScale{0.9}
\Line(20,20)(50,20)
\Line(32,17)(38,23)
\Line(32,23)(38,17)

\end{picture}}
\parbox[c]{12.5cm}{
 \be
= \qquad \bar \psi \gamma_0 \Bigl[ 
-\mu \pm \frac{i}{2}\tilde A a_L + \frac{i}{2}\tilde B (Y_L a_L + Y_R a_R)
\Bigr] \psi,
\ee}

\noindent
where $\mu$ is either $\mu_B/3$ (for quarks) or
$\mu_{L_i}$ (for leptons), $a_L,a_R$ are the left and
right projectors, $Y_{L,R}$ 
are the corresponding hypercharges, and
$\tilde A \equiv gA_0^3, \tilde B \equiv g' B_0$.

\paragraph{Baryon asymmetry.}

In order to account for flavour dependent contributions
to the baryon asymmetry, we have to supplement the effective
potential computed in~\cite{ks96} with the dominant terms
differentiating between the $\mu_{L_i}$'s. Such terms 
must involve leptonic Yukawa couplings, and can arise
either as mass corrections in the 1-loop effective 
potential, or as 2-loop corrections directly 
proportional to the Yukawa couplings~\footnote{At the order
we are working, quark Yukawa couplings are always
associated with terms involving $\mu_B$, and thus do
not make a distinction between the 
different generations.}. For $\phi\sim T$
both types of terms ($\sim m_i^2/T^2,h_i^2$) are of the 
same order of magnitude and must be included simultaneously. 

The dominant fermionic 1-loop mass corrections come from the graph
\vspace*{0.1cm}

\hspace*{5.75cm}
\parbox[c]{2.4cm}
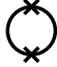
{
\begin{picture}(140,30)(0,0)

\SetWidth{1.5}
\SetScale{0.9}
\CArc(35,15)(10,0,360)
\Line(32,2)(38,8)
\Line(32,8)(38,2)
\Line(32,22)(38,28)
\Line(32,28)(38,22)

\end{picture}}
\parbox[c]{6.25cm}{ }

\vspace*{0.1cm}

\noindent
On the other hand, 
denoting the Higgs with a dashed line, the 2-loop 
graphs potentially contributing are 
(we need only terms with at least one power of $\mu_{L_i}$)

\vspace*{0.1cm}

\hspace*{3.5cm}
\parbox[c]{2.4cm}
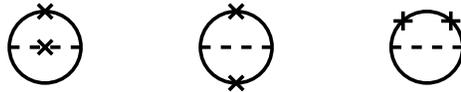
{
\begin{picture}(140,40)(0,0)

\SetWidth{1.5}
\SetScale{0.9}
\CArc(30,20)(15,0,360)
\DashLine(15,20)(45,20){5}
\Line(27,17)(33,23)
\Line(27,23)(33,17)
\Line(27,32)(33,38)
\Line(27,38)(33,32)

\end{picture}}
\parbox[c]{2.4cm}{
\begin{picture}(140,40)(0,0)

\SetWidth{1.5}
\SetScale{0.9}
\CArc(30,20)(15,0,360)
\DashLine(15,20)(45,20){5}
\Line(27,2)(33,8)
\Line(27,8)(33,2)
\Line(27,32)(33,38)
\Line(27,38)(33,32)

\end{picture}}
\parbox[c]{2.4cm}{
\begin{picture}(140,40)(0,0)

\SetWidth{1.5}
\SetScale{0.9}
\CArc(30,20)(15,0,360)
\DashLine(15,20)(45,20){5}
\Line(16,31)(24,31)
\Line(20,27)(20,35)
\Line(36,31)(44,31)
\Line(40,27)(40,35)

\end{picture}}
\parbox[c]{3.0cm}{ }

\vspace*{0.1cm}

\noindent
However, it is easy to see that the first 2-loop graph does in fact  not
contribute. The reason is that viewed as a self-energy  correction
for the Higgs field, the fermionic loop  does not have a term linear
in $\mu$, because the  insertions of $\gamma_0 \mu$ to the two
different  fermion lines cancel each other; thus the result is
$\propto \tilde A^2,\tilde B^2$. 

Furthermore, it turns out the second 2-loop graph does  not contribute
either. It is proportional to the divergent integral 
\be
I = \Tint{P,Q} \frac{p_f(p_f+q_f)}{(P+Q)^2 Q^2 (P^2)^2},
\ee
where $P=(p_f,\vec{p})$ and $p_f$ are the fermionic Matsubara
frequencies. It turns out that the coefficient of this integral  gets
contributions also from the last graph, and altogether the result is
again just a higher order correction of the form $\tilde A^2,\tilde
B^2$.

The remaining finite contribution from the last 2-loop graph, 
together with the 1-loop contribution
as well as all the bosonic contributions, 
give the effective potential 
\ba
V & = & \fr12 \tilde A^2 \biggl[ 
\fr14 \phi^2 +\biggl(
\fr23 + \frac{n_S}{6} + \frac{n_F}{3}
\biggr)T^2
\biggr] + \fr12 \tilde B^2 \biggl[ 
\fr14 \phi^2 +\biggl(
\frac{n_S}{6} + \frac{5 n_F}{9}
\biggr)T^2 
\biggr] + \fr14 \tilde A \tilde B \phi^2 \nn
& + & \frac{i}{3} \tilde B T^2 \biggl(
\frac{n_F}{3}\mu_B -\sum_i \mu_{L_i}
\biggr) - \frac{n_F}{9} \mu_B^2 T^2 - 
\fr14\sum_i \mu_{L_i}^2 T^2 \nn
& - & 
\frac{i}{8\pi^2} \tilde A 
\sum_i \mu_{L_i} m_i^2  +
\frac{T^2}{8\pi^2} 
\sum_i (3i\tilde B\mu_{L_i} + 2 \mu_{L_i}^2) 
\biggl(\frac{m_i^2}{T^2} + \frac{h_i^2}{4} \biggr). \la{bro}
\ea
Working as outlined above, we obtain for the 
physical case ($n_F=3,n_S=1$)
\ba
B & = & 4 \frac{77 T^2+27 \phi^2}{869 T^2+333\phi^2}\sum_i X_i \nn
& + &  
\frac{11}{2\pi^2} 
\frac{47 T^2+18 \phi^2}{869 T^2+333\phi^2}\sum_i X_i \frac{m_i^2}{T^2}
+ \frac{1}{16\pi^2} \frac{1034 T^2+405 \phi^2}{869 T^2+333\phi^2} 
\sum_i X_i h_i^2. \la{final}
\ea
The first term coincides with the one found in~\cite{ks96}. The
latter line is the dominant one if $\sum_i X_i=0$. Using the 
notation of \eq\nr{form}
we have $f_1=(4/13 \pi^2)(143/148) \approx (4/13 \pi^2)\cdot 0.9662$ and
$f_2=(1/13 \pi^2)(585/592) \approx (1/13 \pi^2)\cdot 0.9882$ for $\phi \gg T$,
and $f_1=(4/13 \pi^2)(611/632) \approx (4/13 \pi^2)\cdot 0.9668$ and 
$f_2= f_1/4$ for $\phi\ll T$, in numerical
agreement with the estimates in~\cite{ks88}. 
In the symmetric phase $\phi\ll T$, the analytic result 
for $f_2$ agrees with that given in~\cite{dko}.

When higher
order corrections are taken into account, one expects
$f_i \to f_i [1+{\cal O}(\alpha_W/\pi,\alpha_S/\pi,h^2/\pi^2,m^2/(\pi T)^2)]$, 
where $h$ is a general Yukawa coupling and $m$ a general mass.

\vspace*{0.5cm}

\paragraph{In summary,} we have derived the leading order expressions
for the flavour dependent contributions to the baryon 
asymmetry remaining after a period of full thermal equilibrium,
\eq\nr{final}.
The result should be evaluated at $\phi\sim T$ if the sphaleron 
processes fall out of thermal equilibrium after the system has smoothly
passed from the symmetric to the Higgs phase. 
If, on the contrary, 
there is a strong first order
electroweak phase transition such that the sphaleron processes
are always switched off in the Higgs phase (but at the same 
time no new $B+L$ asymmetry is generated during the transition
due to, say, too little CP-violation),
it should be evaluated at $\phi= 0$. 

\vspace*{0.5cm}

The work of  M.L. was partly supported by the TMR network {\em Finite
Temperature Phase Transitions in Particle Physics}, EU contract no.\
FMRX-CT97-0122. The work of M.S. was partly supported by the Swiss
Science Foundation, contract no. 21-55560.98.

\end{document}